\documentclass{elsart}
\usepackage{amssymb,natbib,amsmath,graphicx,color,subfigure,url,lineno}
\usepackage{epstopdf}
\journal{International Journal of Forecasting}%
\newcommand{\be}{\begin{equation}}
\newcommand{\ee}{\end{equation}}
\newcommand{\bea}{\begin{eqnarray}}
\newcommand{\eea}{\end{eqnarray}}

\begin{document}

%\graphicspath{{Figures/}}

\runauthor{Zhou and Sornette} \markboth{A}{B}

\begin{frontmatter}

\title{Analysis of the real estate market in Las Vegas: Bubble, seasonal
patterns, and prediction of the CSW indexes\thanksref{grants}}

\author[ecust]{{Wei-Xing Zhou}},
\ead{wxzhou@ecust.edu.cn} %
\author[ETH]{{Didier Sornette}\corauthref{cor}}
\corauth[cor]{Corresponding author. Address: KPL F 38.2, Kreuzplatz
5, ETH Zurich, CH-8032 Zurich, Switzerland. Phone: +41 44 632 89 17,
Fax: +41 44 632 19 14.}
\ead{sornette@ethz.ch}%
\ead[url]{http://www.er.ethz.ch/}%

\thanks[grants]{We are grateful to Signature Homes (Las Vegas) for
providing us the data. This work was partially supported by the NSFC
(Grant 70501011), the Fok Ying Tong Education Foundation (Grant
101086), and the Alfred Kastler Foundation which supported W.-X.
Zhou for a visiting position in France.}

\address[ecust]{School of Business, School of Science, and Research
Center of Systems Engineering, East China University of Science and
Technology, Shanghai 200237, China}
\address[ETH]{D-MTEC, ETH Zurich, CH-8032 Zurich, Switzerland}

\begin{abstract}
We analyze 27 house price indexes of Las Vegas from Jun. 1983 to
Mar. 2005, corresponding to 27 different zip codes. These analyses
confirm the existence of a real-estate bubble, defined as a price
acceleration faster than exponential, which is found however to be
confined to a rather limited time interval in the recent past from
approximately 2003 to mid-2004 and has progressively transformed
into a more normal growth rate comparable to pre-bubble levels in
2005. There has been no bubble till 2002 except for a medium-sized
surge in 1990. In addition, we have identified a strong yearly
periodicity which provides a good potential for fine-tuned
prediction from month to month. A monthly monitoring using a model
that we have developed could confirm, by testing the intra-year
structure, if indeed the market has returned to ``normal'' or if
more turbulence is expected ahead. We predict the evolution of the
indexes one year ahead, which is validated with new data up to Sep.
2006. The present analysis demonstrates the existence of very
significant variations at the local scale, in the sense that the
bubble in Las Vegas seems to have preceded the more global USA
bubble and has ended approximately two years earlier (mid 2004 for
Las Vegas compared with mid-2006 for the whole of the USA).
\end{abstract}


\end{frontmatter}

%\tableofcontents

%\linenumbers

\section{Introduction}
\label{sec:intro}

\citet{Zhou-Sornette-2003a-PA} analyzed the deflated quarterly
average sales prices $p(t)$ from December 1992 to December 2002 of
new houses sold in all the states in the USA and by regions
(northeast, midwest, south and west) and found that, while there was
undoubtedly a strong growth rate, there was no evidence of a bubble
in the latest six years (as qualified by a super-exponential
growth). Then, \citet{Zhou-Sornette-2006b-PA} analyzed the quarterly
average sale prices of new houses sold in the USA as a whole, in the
northeast, midwest, south, and west of the USA, in each of the 50
states and the District of Columbia of the USA up to the first
quarter of 2005, to determine whether they have grown
faster-than-exponential (which is taken as the diagnostic of a
bubble). \citet{Zhou-Sornette-2006b-PA} found that 22 states (mostly
Northeast and West) exhibit clear-cut signatures of a fast growing
bubble. From the analysis of the S\&P 500 Home Index, they concluded
that the turning point of the bubble would probably occur around
mid-2006. The specific statement found at the bottom of page 306 of
Ref.[\cite{Zhou-Sornette-2006b-PA}] is: ``We observe a good
stability of the predicted $t_c \approx$  mid-2006 for the two LPPL
models (2) and (3). The spread of $t_c$ is larger for the
second-order LPPL fits but brackets mid-2006. As mentioned before,
the power-law fits are not reliable. We conclude that the turning
point of the bubble will probably occur around mid-2006.'' It should be
stressed that these studies departed from most other
reports by analysts and consulting firms on real estate prices in
that \citet{Zhou-Sornette-2003a-PA,Zhou-Sornette-2006b-PA} did
{\em{not}} characterize the housing market as overpriced in 2003. It
is only in 2004-2005 that they confirmed that the signatures of an
unsustainable bubble path has been revealed.

\begin{figure}[htb]
\begin{center}
\includegraphics[width=7.5cm]{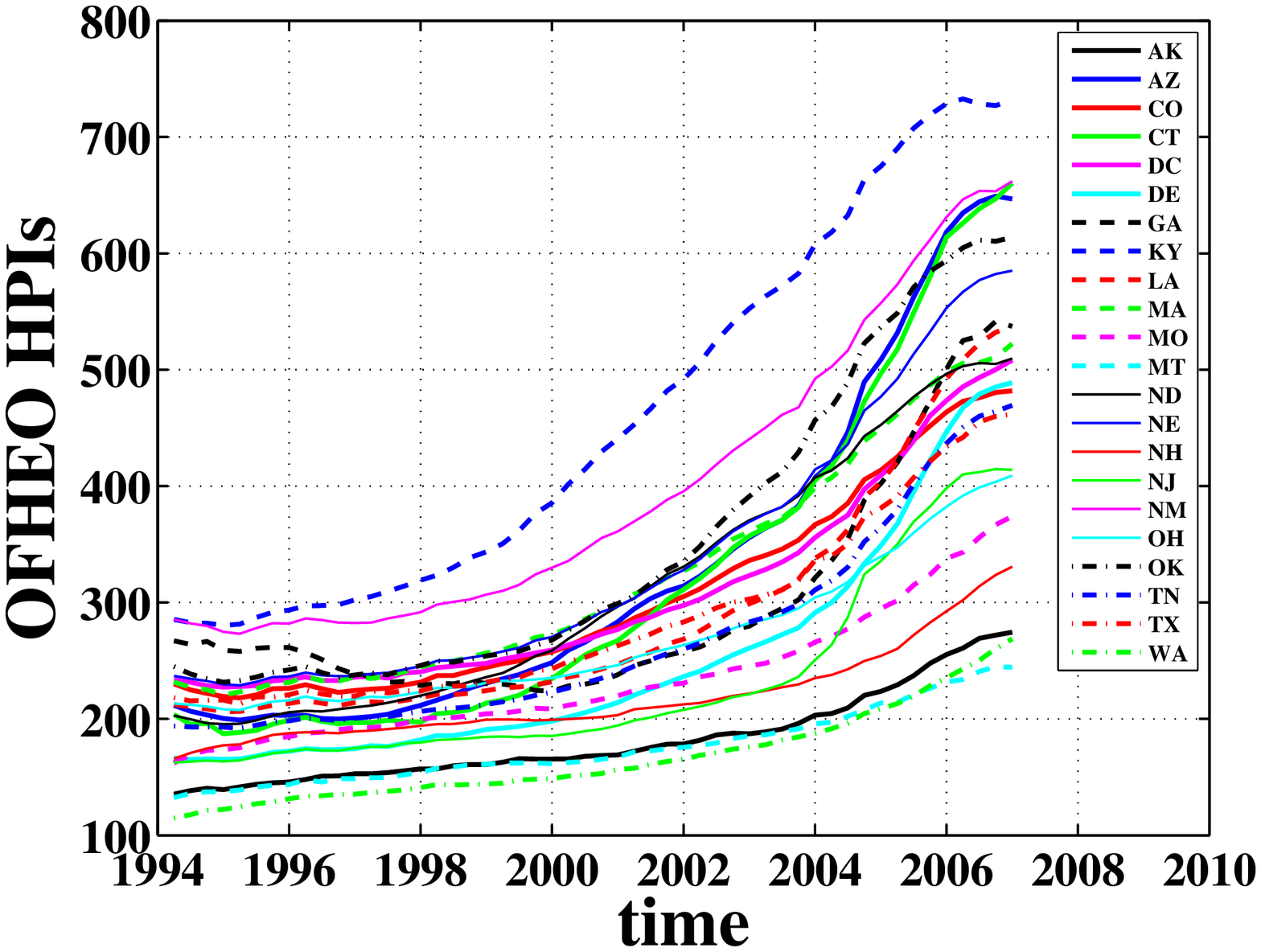}
\includegraphics[width=7.5cm]{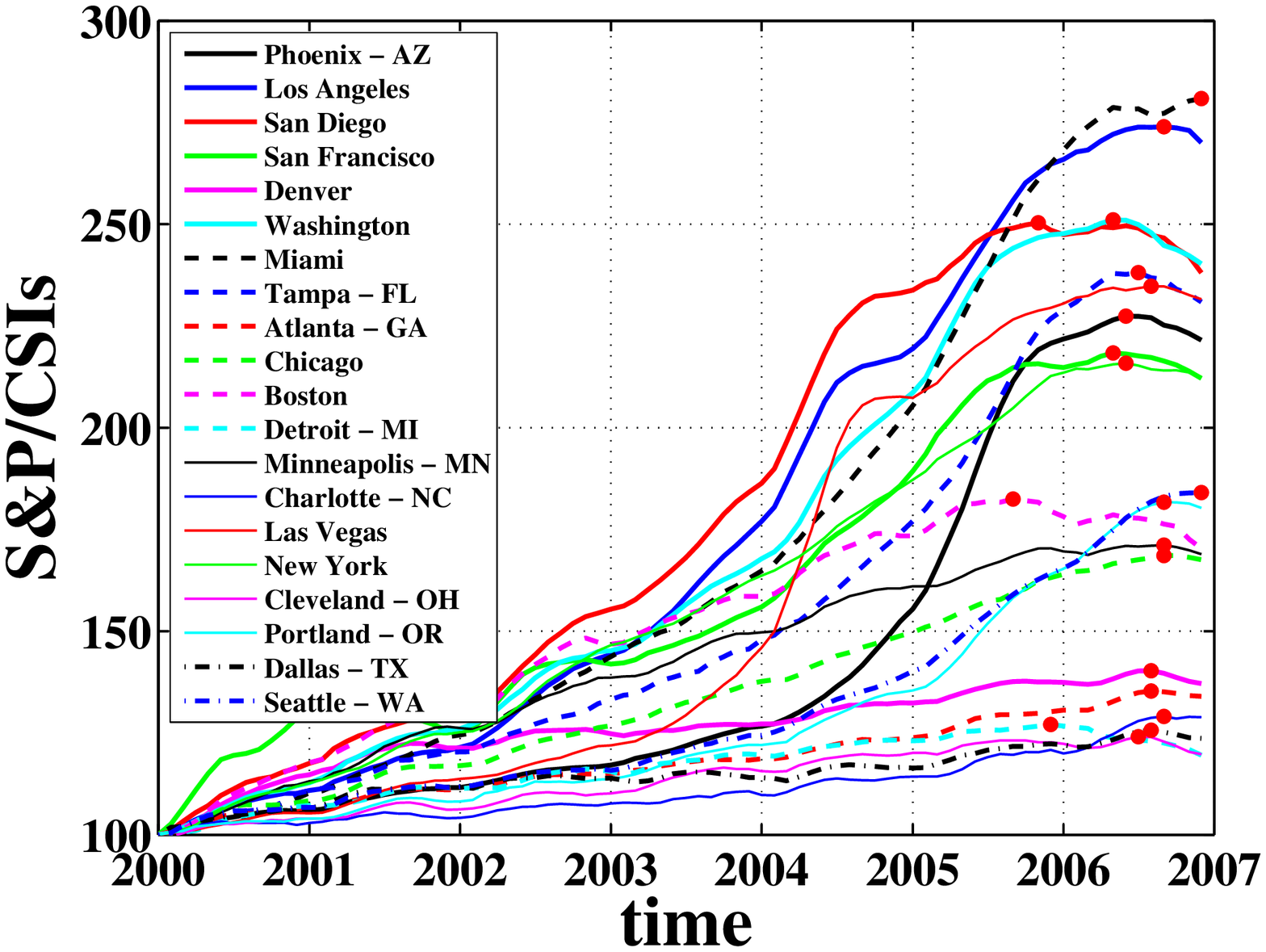}
\end{center}
\caption{Evaluation of the prediction of
\citet{Zhou-Sornette-2006b-PA} that ``the turning point of the bubble
will probably occur around mid-2006'' using the OFHEO HPI data (upper
panel) and the S\&P CSI data (lower panel).}
\label{Fig:US:house:Pred:Reevaluate}
\end{figure}

Let us briefly analyze how this prediction has fared. The upper
panel of Figure \ref{Fig:US:house:Pred:Reevaluate} shows the
quarterly house price indexes (HPIs) in the 21 states and in the
District of Columbia (DC) from 1994 to the fourth quarter of 2006
released by the OFHEO. It is evident that the growth in most of
these 22 HPIs has slowed down or even stopped during the year of
2006. When we look at the S\&P Case-Shiller Home Indexes of the 20
major US cities, as illustrated in the lower panel of Figure
\ref{Fig:US:house:Pred:Reevaluate}, we observe that the majority of
the S\&P/CSIs had a maximum denoted by a solid dot in the middle of
2006, validating the prediction of \citet{Zhou-Sornette-2006b-PA}.
Specifically, the times of the maxima are respectively 2006/06/01,
2006/09/01, 2005/11/01, 2006/05/01, 2006/08/01, 2006/05/01,
2006/12/01, 2006/07/01, 2006/08/01, 2006/09/01, 2005/09/01,
2005/12/01, 2006/09/01, 2006/09/01, 2006/08/01, 2006/06/01,
2006/07/01, 2006/09/01, 2006/08/01, 2006/12/01, 2006/06/01, and
2006/07/01 for the 20 cities shown in the legend of the lower panel.
The only two cities with a maximum occurring later towards the end
of 2006 (2006/12/01) are Miami and Seattle. However their growth
rates decreased remarkably in 2006 as shown in the figure.
Furthermore, the S\&P/CS Home Price Composite-10 reached its
historical high 226.29 on 2006/06/01 and the Composite-20 culminated
to 206.53 on 2006/07/01, again confirming remarkably well the
validity of the forecast of \cite{Zhou-Sornette-2006b-PA}.

In this note, we provide a more regional study of the diagnostic of
bubbles and the prediction of their demise. Specifically, we analyze
the Case-Shiller-Weiss (CSW) Zip Code Indexes of 27 different Las
Vegas regions calculated with a monthly rate from June-1983 to
March-2005. The CSW Indexes are based on the so-called repeat sales
methods which directly measure house price appreciations. The key to
these data is that they are observations of multiple transactions on
the same property, repeated over many properties and then pooled in
an index. Prices from different time periods are combined to create
``matched pairs,'' providing a direct measure of price changes for a
given property over a known period of time.
\citet{Bailey-Muth-Nourse-1963-JASA} proposed the basic repeat sales
method over four decades ago, but only after the work by
\citet{Case-Shiller-1987-NEER,Case-Shiller-1989-AER,Case-Shiller-1990-AREUEAJ}
did the idea receive significant attention in the housing research
community.

Studying the Las Vegas database is particular suitable since Las
Vegas belongs to a state which was identified by
\citet{Zhou-Sornette-2006b-PA} as one of the 22 states with a fast
growing bubble in 2005. With access to 27 different CSW Zip Code
Indexes of Las Vegas, we are able to obtain more reliable and
fine-grained measures, which both confirm and extend the previous
analyses of \citet{Zhou-Sornette-2003a-PA,Zhou-Sornette-2006b-PA}.
The next section recalls the conceptual background underlying our
empirical approach. Then, section \ref{s1:Regimes} analyzes the
regional CSW indexes for Las Vegas, showing that there is a regime
shift separated by a bubble around year 2004. Section
\ref{s1:periodic} identifies and then analyzes the yearly
periodicity and intra-year pattern detected in the growth rate of
the regional CSW indexes. Section \ref{s1:prediction} offers a
preliminary forecast based on the periodicity analyses in
Sec.~\ref{s1:periodic}. Section~\ref{s1:concl} concludes.

\section{Conceptual background of our empirical analysis}
\subsection{Humans as social animals and herding}

Humans are perhaps the most social mammals and they shape
their environment to their personal and social needs. This statement
is based on a growing body of research at the frontier between new
disciplines called neuro-economics, evolutionary psychology,
cognitive science, and behavioral finance. This body of evidence
emphasizes the very human nature of humans with its biases and
limitations, opposed to the previously prevailing view of rational
economic agents optimizing their decisions based on unlimited access
to information and to computation resources.

Here, we focus on an empirical question (the existence and detection of
real-estate bubbles) which, we hypothesize,
is a footprint of perhaps the most robust trait of
humans and the most visible imprint in our social
affairs: imitation and herding. Imitation has been documented in
psychology and in neuro-sciences as one of the most evolved
cognitive process, requiring a developed cortex and sophisticated
processing abilities. In short, we learn our basics and how to adapt
mostly by imitation all along our life. It seems that imitation has
evolved as an evolutionary advantageous trait, and may even have
promoted the development of our anomalously large brain (compared
with other mammals). It is actually ``rational'' to imitate when
lacking sufficient time, energy and information to take a decision
based only on private information and processing, that is..., most
of the time. Imitation, in obvious or subtle forms, is a pervasive
activity of humans. In the modern business, economic and financial
worlds, the tendency for humans to imitate leads in its strongest
form to herding and to crowd effects.

Based on a theory of cooperative herding and imitation, we have
shown that imitation leads to positive feedbacks, that is, an action
leads to consequences which themselves reinforce the action and so
on, leading to virtuous or vicious circles. We have formalized these
ideas in a general mathematical theory which has led to observable
signature of herding, in the form of so-called log-periodic power
law acceleration of prices. A power law acceleration of prices
reflects the positive feedback mechanism. When present,
log-periodicity takes into account the competition between positive
feedback (self-fulfilling sentiment), negative feedbacks
(contrariant behavior and fundamental/value analysis) and inertia
(everything takes time to adjust). \citet{Sornette-2003} presented a
general introduction, a synthesis and examples of applications.

\subsection{Definition and mechanism for bubbles}

The term ``bubble'' is widely used but rarely clearly defined.
Following \citet{Case-Shiller-2003-BPEA}, the term ``bubble'' refers
to a situation in which excessive public expectations of future
price increases cause prices to be temporarily elevated. During a
housing price bubble, homebuyers think that a home that they would
normally consider too expensive for them is now an acceptable
purchase because they will be compensated by significant further
price increases. They will not need to save as much as they
otherwise might, because they expect the increased value of their
home to do the saving for them. First-time homebuyers may also worry
during a housing bubble that if they do not buy now, they will not
be able to afford a home later. Furthermore, the expectation of
large price increases may have a strong impact on demand if people
think that home prices are very unlikely to fall, and certainly not
likely to fall for long, so that there is little perceived risk
associated with an investment in a home.

What is the origin of bubbles? In a nutshell, speculative bubbles
are caused by ``precipitating factors'' that change public opinion
about markets or that have an immediate impact on demand, and by
``amplification mechanisms'' that take the form of price-to-price
feedback, as stressed by \citet{Shiller-2000}. A number of
fundamental factors can influence price movements in housing
markets. On the demand side, demographics, income growth, employment
growth, changes in financing mechanisms or interest rates, as well
as changes in location characteristics such as accessibility,
schools, or crime, to name a few, have been shown to have effects.
On the supply side, attention has been paid to construction costs,
the age of the housing stock, and the industrial organization of the
housing market. The elasticity of supply has been shown to be a key
factor in the cyclical behavior of home prices. The cyclical process
that we observed in the 1980s in those cities experiencing
boom-and-bust cycles was caused by the general economic expansion,
best proxied by employment gains, which drove demand up. In the
short run, those increases in demand encountered an inelastic supply
of housing and developable land, inventories of for-sale properties
shrank, and vacancy declined. As a consequence, prices accelerated.
This provided an amplification mechanism as it led buyers to
anticipate further gains, and the bubble was born. Once prices
overshoot or supply catches up, inventories begin to rise, time on
the market increases, vacancy rises, and price increases slow down,
eventually encountering downward stickiness. The predominant story
about home prices is always the prices themselves
\citep[see][]{Shiller-2000,Sornette-2003}; the feedback from initial
price increases to further price increases is a mechanism that
amplifies the effects of the precipitating factors. If prices are
going up rapidly, there is much word-of-mouth communication, a
hallmark of a bubble. The word of mouth can spread optimistic
stories and thus help cause an overreaction to other stories, such
as stories about employment. The amplification can also work on the
downside as well. Price decreases will generate publicity for
negative stories about the city, but downward stickiness is
encountered initially.

\subsection{Was there a bubble? Status of the argument based
on the ratio of cost of owning versus cost of renting}

In recent years, there has been increasing debates on whether there
was a real estate bubble or not in the United States of America.
\citet{Case-Shiller-2003-BPEA}, \cite{Shiller-2006-EV} and
\citet{Smith-Smith-2006-BPEA} argued that the house prices over the
period 2000-2005 were not abnormal as they reflected only the
convergence of the prices to their fundamentals from below. In
contrast, \citet{Zhou-Sornette-2006b-PA} and
\citet{Roehner-2006-EIER} have suggested that there was a bubble,
which became identifiable only after 2003, that is,  after the work of
\citet{Zhou-Sornette-2003a-PA}.

In this context, it is instructive to comment on the study by
\citet{Himmelberg-Mayer-Sinai-2005}, from the Federal Reserve Bank
of New York , as it reflects the never ending debate between tenants
of the fundamental valuation explanation and those invoking
speculative bubbles. We are resolutely part of the second group.
\citet{Himmelberg-Mayer-Sinai-2005} constructed measures of the
annual cost of single-family housing for 46 metropolitan areas in
the United States over the last 25 years and compared them with
local rents and incomes as a way of judging the level of housing
prices. In a nutshell, they claimed in 2005 that conventional
metrics like the growth rate of house prices, the price-to-rent
ratio, and the price-to-income ratio can be misleading and lead to
incorrect conclusions on the existence of the real-estate bubble.
Their measure showed that, during the 1980s, houses looked most
overvalued in many of the same cities that subsequently experienced
the largest house price declines. But they found that from the
trough of 1995 to 2004, the cost of owning rose somewhat relative to
the cost of renting, but not, in most cities, to levels that made
houses look overvalued.

The rosy conclusion of \citet{Himmelberg-Mayer-Sinai-2005}, that
2004-2005 prices were justifiable and that there was no risk of
deflation as no bubble was present, is based on a particularly
curious comparison between cost of owning and cost of renting, as
noticed by \citet{Jorion-2005-WSJ}, in a letter to the Wall Street
Journal. Indeed, they candidly revealed however that their ``cost of
owning'' calculations imply an ``expected appreciation on the
property'' coefficient. The value for this factor is no doubt
derived from figures for appreciation as currently observed on the
housing market, meaning they regarded the current appreciation level
as a reasonable assumption for what would indeed happen next --
which is precisely what our analyses and that of others question. In
other words, the authors had unwittingly hard-wired into their model
the assertion that there was no housing bubble; little wonder then
that this is also what they felt authorized to conclude. The
circularity of their reasoning is particularly obvious in an
illustration they gave for San Francisco where for more than 60
years the price-to-rent ratio has exceeded the national average,
which, so they claimed, ``does not necessarily make owning there
more expensive than renting.'' The reason why is that ``high
financing costs are offset by above-average expected capital
gains.'' Translated, this means that as long as there is a bubble,
prices will go up and investing in a house remain a profitable
operation. This trivial statement is hollow; the real question is
whether the trend that is observed now remains sustainable.

In addition to this criticism put forward by
\citet{Jorion-2005-WSJ}, there are other reasons to doubt the
validity of the conclusion of \citet{Himmelberg-Mayer-Sinai-2005}.
In the own words of \citet{Himmelberg-Mayer-Sinai-2005}, ``the ratio
of the cost of owning to the cost of renting is especially sensitive
to the real long-term interest rates.'' They are right in their rosy
conclusion... as long as the long-term interest rates remain
exceptionally low. It is particularly surprising that their
estimation of the ratio of the cost of owning to the cost of renting
was based on the most recent rates over the preceding year of their
analysis (2004), while the price of a house is a long-term
investment: what will be the long-term rates in 10, 20, 30, or 50
years? Another problem is that their analysis was
``mono-dimensional'': they proposed that everything depends only on
the ratio of the cost of owning to the cost of renting. But they
missed the interest rates as an independent variable. As a
consequence, it is not reasonable to compare the 1980s and the
present time, as the long-term interest rates had nothing in common.
Another problem with their analysis is that they assumed
``equilibrium,'' while people are sensitive to the history-dependent
path followed by the prices. In other words, people are sensitive to
the way prices reach a certain level, if there is an acceleration
that can self-fuel itself for a while, while
\citet{Himmelberg-Mayer-Sinai-2005} discussed only the
mono-dimensional level of the price, and not how it got there. We
think that this general error made by ``equilibrium'' economists
constitutes a fundamental flaw which fails to capture the real
nature of the organization of human societies and their decision
process. In the sequel, we actually focus our attention on
signatures of price trajectories that highlight the importance of
history dependence for prediction.

This discussion is reminiscent of the proposition by
\citet{Mauboussin}, offered close to the peak of the Internet and new
technology bubble that culminated in 2000, that better business
models, the network effect, first-to-scale advantages, and real
options effect could account rationally for the high prices of
dot.com and other New Economy companies. These interesting views
expounded in early 1999 were in synchrony with the bull market of
1999 and preceding years. They participated in the general
optimistic view and added to the strength of the herd. Later, after
the collapse of the bubble, these explanations seem less attractive.
This did not escape the then U.S. Federal Reserve chairman Alan
\citet{Greenspan97}, who said : ``Is it possible that there is
something fundamentally new about this current period that would
warrant such complacency? Yes, it is possible. Markets may have
become more efficient, competition is more global, and information
technology has doubtless enhanced the stability of business
operations. But, regrettably, history is strewn with visions of such
new eras that, in the end, have proven to be a mirage. In short,
history counsels caution.''

\section{Regime shift in the CSW Zip Code Indexes of Las Vegas}
\label{s1:Regimes}

\subsection{Description of the data}

\begin{figure}[htb]
\begin{center}
\includegraphics[width=8cm]{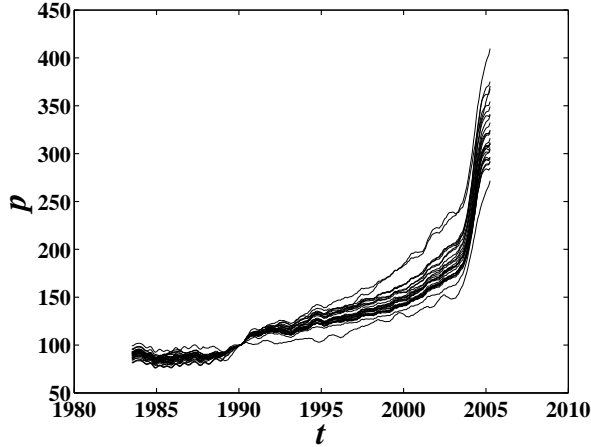}
\end{center}
\caption{Time evolution of the Case-Shiller-Weiss (CSW) Zip Code
Indexes of 27 Las Vegas zip regions from June-1983 to March-2005.}
\label{Fig:LV_CSW_pt}
\end{figure}

We now turn to the analysis of the CSW indexes of 27 different Las
Vegas zip regions obtained with a monthly rate. The 27 monthly CSW
data sets start from June-1983 and end in March-2005. Figure
\ref{Fig:LV_CSW_pt} shows the price trajectories of all the 27 CSW
indexes. Visual inspection shows (i) a very similar behavior of all
the different zip codes and (ii) a sudden increase of the indexes
since Mid-2003. Let us now analyze this data quantitatively.

\subsection{Power law fits}

The simplest mathematical equation capturing the positive feedback
effect and herding is the power law formula \citep[see][for a simple
introduction in a similar context]{Broekstra-Sornette-Zhou-2005-PA}
\begin{equation}
I(t) = A+B |t_c-t|^m~ ~,\label{Eq:PL}
\end{equation}
where $B<0$ and $0 < m<1$ or $B>0$ and $m<1$. Others cases do not
qualify as a power law acceleration. For $B<0$ and $0 < m<1$ or
$B>0$ and $m<0$, the trajectory of $I(t)$ described by (\ref{Eq:PL})
expresses the existence of an accelerating bubble, which is faster
than exponential. This is taken as one hallmark of the existence of
a bubble.

Notice also that this formula expresses the existence of a
singularity at time $t_c$, which should be interpreted as a change
of regime (the mathematical singularity does not exist in reality
and is rounded off by so-called finite-size effects and the
appearance of a large susceptibility to other mechanisms). This
critical time $t_c$ must be interpreted as the end of the bubble and
the time where the regime is transiting to another state through a
crash or simply a plateau or a slowly moving correction.

\begin{figure}[htb]
\begin{center}
\includegraphics[width=8cm]{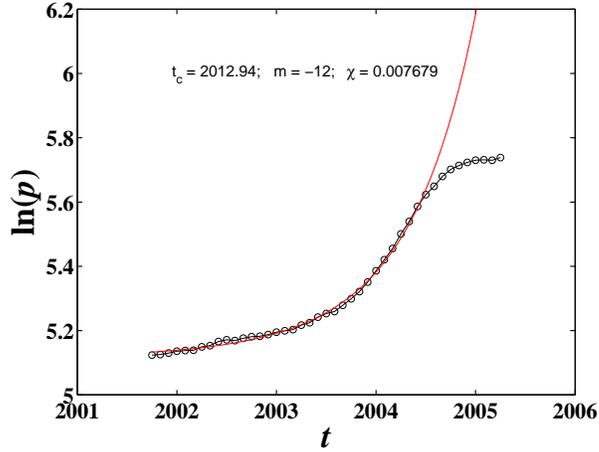}
\end{center}
\caption{Typical evolution of a CSW index from Aug-2001 to Jun-2005
and its fit by a power law, showing both the faster-than-exponential
growth up to mid-2004 and the smooth transition to a much slower
growth at later times. The  root-mean-square  $\chi$
of the residuals of the fit as well at $t_c$ and $m$ are given inside the figure.} \label{FigLV_Pred_CSW_PLs_Z89120S_short}
\end{figure}

We have fitted each of the 27 individual CSW indexes using the pure
power law model (\ref{Eq:PL}). The data used for fitting is from
Dec-1995 to Jun-2005. We do not show the results as the signature of
a power law growth is not evident, essentially because the
acceleration is only over a rather short period of time from
approximately 2002 to 2004. As a consequence, power law fits give
unreliable critical time $t_c$ too much in the future (like 2008 and
beyond). We have thus redone the fits of the 27 CSW indexes over a
shorter time interval from Aug-2001 to Jun-2005. A typical example
is shown in Fig~\ref{FigLV_Pred_CSW_PLs_Z89120S_short}. All other 26
CSW are very similar, with some variations of the parameters, but
the message is the same: while there is a clear
faster-than-exponential acceleration over most of the time interval,
the price trajectory has clearly transitioned into another regime in
the latter part of the time interval considered here. The transition
occurred smoothly from mid-2004 to mid-2005 (the end of the time
period analyzed here).

It is important to recognize that the power law regime is expected
only relatively close to the critical time $t_c$, while other
behaviors are expected far from $t_c$. The simplest model is to
consider that, far from $t_c$, the price follows an exponential
growth with an approximately constant growth rate $\mu$: \be I(t) =
a + b e^{\mu t}~. \label{mgms} \ee A fuller description is thus to
consider that formula (\ref{mgms}) holds from the beginning of the
time series up to a cross-over time $t^*$, beyond which expression
(\ref{Eq:PL}) takes over. Any given price trajectory should thus be
fitted by (\ref{mgms}) from some initial time $t_{\rm start}$ to
time $t^*$ and then by (\ref{Eq:PL}) from $t^*$ to the end of the
time series. Technically, $t^*$ is known from the parameters $a, b,
\mu, A, B, t_c, m$ by the condition of continuity of $I(t)$ at
$t=t^*$, that is, both formulas give the same value at $t=t^*$. We
can further determine one of the parameters $a, b$ or $\mu$ by
imposing a condition of differentiability at $t^*$, that is, the
first time-derivative of $I(t)$ is continuous at $t^*$. This
approach is known in numerical analysis as ``asymptotic matching''
\citep[see][]{Bender-Orszag-1978}.

A simplified description of such a cross-over between a standard
exponential growth and the power law super-exponential acceleration
is obtained by using a more compact formulation
\begin{equation}
I(t) = A+B\tanh[(t_c-t)/\tau]^m~, \label{Eq:tanh}
\end{equation}
where $\tanh$ denotes the hyperbolic tangent function. This
expression derives from a study of the transition from the
non-critical to critical regime in rupture processes (of which
bubbles and their terminal singularity belong to) conducted by
\citet{Sornette-Andersen-1998-EPJB}. This expression has the virtue
of providing automatically a smooth transition between the
exponential behavior (\ref{mgms}) and the pure power law
(\ref{Eq:PL}), since $\tanh[(t_c-t)/\tau] \approx (t_c-t)/\tau$ for
$t_c-t < \tau$ and $\tanh[(t_c-t)/\tau] \approx 1- 2
e^{2(t-t_c)/\tau}$ for $t_c-t > \tau$. In this later case $t_c-t >
\tau$, expression (\ref{Eq:tanh}) becomes of the form (\ref{mgms})
with $m=1$ and \bea
a &=& A+B~,  \\
b &=& -2B e^{-2t_c/\tau}~, \\
\mu &=& 1/\tau~. \eea In contrast, for $t_c-t < \tau$, expression
(\ref{Eq:tanh}) becomes of the form (\ref{Eq:PL}) with the
correspondence $B/\tau^m \to B$. Expression (\ref{Eq:tanh}) has only
five free parameters, in contrast with the model involving the
cross-over from (\ref{mgms}) to (\ref{Eq:PL}) which has $7$ free
parameters ($a, b, \mu, A, B, t_c, m$) while $t^*$ is determined by
the asymptotic matching). The pure power law formula (\ref{Eq:PL})
has $4$ parameters while the exponential law (\ref{mgms})  has just
$3$ parameters. The problem with expression (\ref{Eq:tanh}) is that
it does not recover a pure exponential growth even for $t_c-t >
\tau$, when $m \neq 1$. Thus, expression (\ref{Eq:tanh}) is limited
in fully describing a possible cross-over from a standard mild
exponential growth and an super-exponential power law acceleration.
Our tests (not shown) find that a fit with model (\ref{Eq:tanh})
retrieve the pure power law model (\ref{Eq:PL}) with the same
critical time $t_c$ and exponent $m$ and the same root-mean-square
residual r.m.s. (the fit adjusts the parameter $\tau$ to a very
large value, ensuring that the fit is always in the regime $t_c-t
\ll \tau$ so that the hyperbolic tangential model reduces to the
pure power law model). Thus, contrary to our initial hopes, this
approach does not provide any additional insight.

Inspired by these tests, we could propose the following modified
model \be I(t) = a + b e^{\mu t} (t_c-t)^m~. \label{exppower} \ee It
has $5$ adjustable parameters, like model (\ref{Eq:tanh}), but it
seems more flexible to describe the looked-for cross-over: for large
$t_c-t$, the power law term $(t_c-t)^m$ changes slowly, especially
for $0<m<1$ as is expected here; for small $t_c-t$, the power law
term changes a lot while the exponential term is basically constant.
But, this model is correct for a critical point only if $m<0$ so
that $b>0$; otherwise, if $0<m<1$, $b<0$ and for $t_c-t$ large, the
exponential term which dominate does not describe a growth but an
exponentially accelerating decay. For $0<m<1$, we thus need a
different formulation. We propose \be I(t) = a + b e^{\mu t} + c
(t_c-t)^m~. \label{exppower2} \ee We have fitted this formula to the
data over the four periods 1983 - Oct. 2004, 1991 - Oct. 2004, 1983
- Mar. 2005, 1991 - Mar. 2005 and, while the fits are reasonable,
the critical time $t_c$ is found to overshoot to 2007-2008, which is
a typical signature that the model is not predictive.

In conclusion of this first preliminary study, the presence of a
bubble (faster-than-exponential growth) is confirmed but the
determination of the end of this phase is for the moment unreliable.

\subsection{Dependence of the growth rate on the index value}

\begin{figure}[htb]
\begin{center}
\includegraphics[width=8cm]{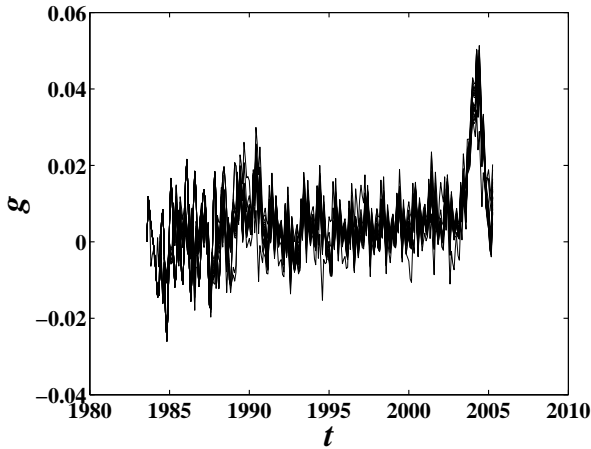}
\end{center}
\caption{Evolution of the growth rates of the 27 regional CSW
indexes from June-1983 to March-2005.} \label{Fig:LV_CSW_gt}
\end{figure}

The monthly growth rate $g(t)$ of a given CSW index at time $t$ is defined by
\begin{equation}
g(t) = \ln[p(t)/p(t-1)]~, \label{eq:g}
\end{equation}
where $p(t)$ is the price of that CSW index at time $t$. Figure
\ref{Fig:LV_CSW_gt} shows the evolution of the growth rates of the
27 CSW indexes from June-1983 to March-2005. While there are some
variations, all 27 CSW indexes follow practically the same pattern.
We clearly observe a large peak of growth over the period 2003-2005.
Notice that this recent peak is much larger and coherent than the
previous one ending in 1991, which was followed by a price
stabilization and even a price drop in certain cases. This figure
stresses that the acceleration in growth rate is a very localized
event which occurred essentially in 2003-2004 and the subsequent
growth rate has leveled off to pre-bubble times. We can conclude
that there has been no bubble from 1990 to 2002, approximately, then
a short-lived bubble until mid-2004 followed by a smoothed
transition back to normal.

\begin{figure}[htb]
\begin{center}
\includegraphics[width=8cm]{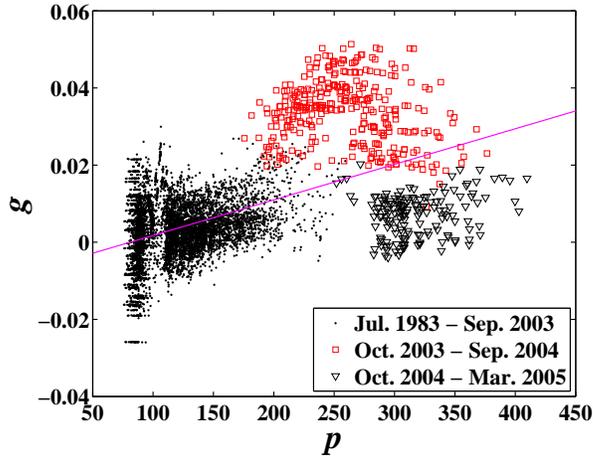}
\end{center}
\caption{Dependence on the data price $p$ for all CSW indexes of its
growth rate $g$. The overall correlation coefficient is 0.494. The
red line is the linear fit of the data points.}
\label{Fig:LV_CSW_pg}
\end{figure}

Fig.~\ref{Fig:LV_CSW_pg} plots the price
growth rate $g(t)$ versus the price $p(t)$ itself for the 27 CSW indexes.
A linear regression of the data points on
Fig.~\ref{Fig:LV_CSW_pg}, shown as the red straight line, gives a
correlation coefficient of 0.494. If we perform linear
regression for each index, then we find an average correlation
coefficient $0.503 \pm 0.036$, confirming the robustness
of this estimation of the correlation between growth rate
and price level. The obtained relation between $g$ and $p$ obtained
from this correlation analysis is captured by the following
mathematical regression
\begin{equation}
g = 0.00922 \times {p \over 100} - 0.00747~. \label{Eq:gp}
\end{equation}
In words, if $p$ is large, then $g$ is large on average, which
confirms the concept of a positive feedback of price on its further
growth. The continuous time
limit of $g(t)$ defined by (\ref{eq:g}) is
\be
g(t) = {d \ln p \over dt} = {1 \over p} {d p \over dt}~.
\ee
This last equation together with (\ref{Eq:gp}), that we
write as $g(t) = \alpha p - \beta$ (with $\alpha = 0.00922/100$ and
$\beta = 0.00747$), implies the following
ordinary differential equation
\be
{dp \over dt} = \alpha p^2 - \beta p~,
\ee
which indeed gives a power law acceleration $p(t) \sim 1/(t_c-t)$
asymptotically close to the critical time $t_c$. Note that this
critical time is determined by the initial conditions, and is called
in mathematics a movable singularity. We conclude from this
first analysis that the rough linear growth of the growth rate
confirms the existence of a bubble growing faster than exponential
according to an approximate power law. But of course, the exponent
of this power law is poorly constrained, in particular from the fact
that the growth rate $g(t)$ exhibits significant variability
and furthermore nonlinearity, as can be seen in Fig.~\ref{Fig:LV_CSW_pg}.

It is useful to refine this analysis by separating the whole time
interval into three distinct intervals. The corresponding plot of
the growth rate $g$ as a function of price is shown in
Fig.~\ref{Fig:LV_CSW_pg} with different symbols: period 1 is Jul.
1983 to Sept. 2003, period 2 is from Oct. 2003 to Sept. 2004, and
period 3 is from Oct. 2004 to Mar. 2005. An anomaly can be clearly
outlined, associated with the red dots which correspond to the
anomalous peak in the growth rate in the period from Oct. 2003 to
Sept. 2004. Notice also that the most recent time interval from Oct.
2004 to Mar. 2005 shows practically the same behavior as the first
period before 2003. In other words, when removing the data in red
for the period from Oct. 2003 to Sept. 2004, the growth rate $g(t)$
is practically independent of $p$, which qualifies the normal
regime. We can thus conclude that this so-called ``phase-portrait''
of the growth rate versus price has identified clearly an anomalous
time interval associated with extremely fast accelerating prices
followed by a more recent period where the price growth has resumed
a more normal regime.

\section{Yearly periodicity and intra-year structure}
\label{s1:periodic}

\subsection{Yearly periodicity from superposed year analysis and
spectral analysis}

\begin{figure}[htb]
\begin{center}
\includegraphics[width=8cm]{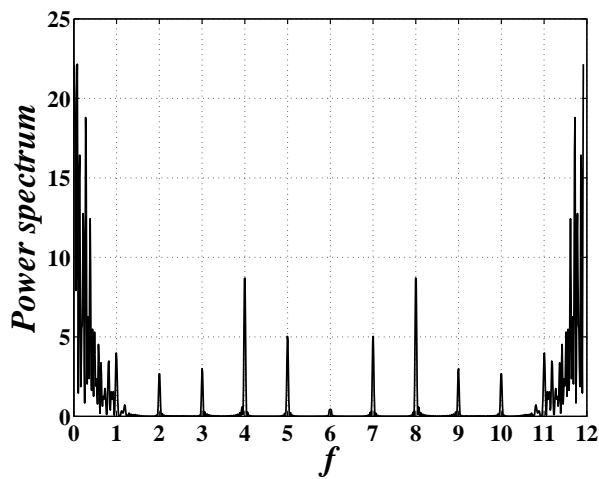}
\end{center}
\caption{Spectrum analysis to confirm the strong periodicity in
$g(t)$.} \label{Fig:LV_Lomb}
\end{figure}

\begin{figure}[htb]
\begin{center}
\includegraphics[width=8cm]{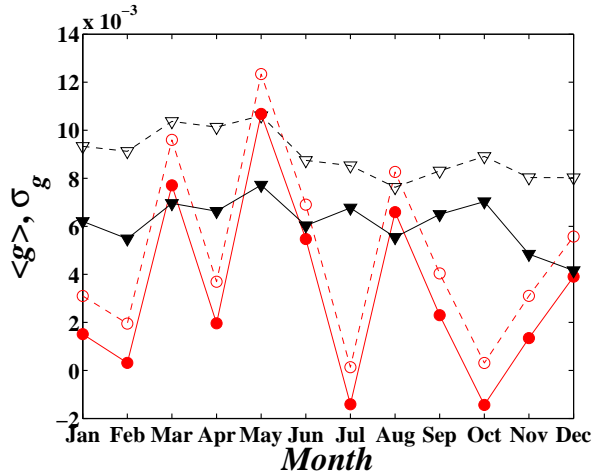}
\end{center}
\caption{Monthly average growth rate (circles) and its standard
deviation (triangles) as a function of the month within the year.
Dash: results obtained over all 27 indexes over the period from Jun.
1983 to Mar. 2005; Solid: results obtained over all 27 indexes over
the period from Jun. 1983 to May 2003.} \label{Fig:LV_Pred_g}
\end{figure}

In Fig.~\ref{Fig:LV_CSW_gt}, the time dependence of the monthly
growth rate exhibits a  clear seasonality (or periodicity), which
appear visually to be predominantly a yearly phenomenon. This visual
observation is made quantitative by performing a spectral Fourier
analysis. The  power spectrum of a typical CSW index is shown in
Fig.~\ref{Fig:LV_Lomb} (all CSW indexes show the same power
spectrum). Since the unit of time used here is one year, the
frequency $f$ is in unit of $1/$year. A periodic behavior with
period one year should translate into a peak at $f=1$ plus all its
harmonics  $f=2, 3, 4, \cdots$, which is indeed observed in
Fig.~\ref{Fig:LV_Lomb}. Note also that the spectrum has large peaks
at $f=4$ and $f=8$ among the harmonics of $f=1$, which indicates a
weak periodicity with period of one quarter. This is consistent with
Fig.~\ref{Fig:LV_Pred_g}, where four oscillations in the averaged
monthly growth rates can be observed.

Note that the power spectrum itself is periodic with a
period of 12, which is the sampling frequency, equal to the double of
the Nyquist
frequency. There are also many peaks in the low-frequency region (larger
that one-year time scale) close to
$f=0$, which are associated with the time scales
of the global trends produced by the big peaks in
$g(t)$ around year 2004 as well as around 1990.

To further explore this seasonal variability of the price growth
rates, we calculate the averages of the growth rates for given
months, where the average is performed over all years. Consider for
instance the month of January: we look up the growth rate for all
the data over all years for the month of January and take the
average. We do the same for each successive months. The result is
shown in Figure \ref{Fig:LV_Pred_g} for two time periods, which
gives the average growth rate $\langle{g}\rangle$ for different
months of the year. The red dash line and circles give the resultant
$\langle{g}\rangle$ for all the data and the black dash line and
triangles give the standard deviation $\sigma_g$ for all data (which
is a measure of the variability from year to year and from zip code
to zip code around the average). The difference between the two time
periods is precisely the time interval from June 2003 to March 2005:
this period is responsible for a significant increase of the average
growth rate (compare the red dashed line (filled circles) with the
red continuous line (open circles)) and an even larger increase of
the variability (compare the black continuous line (filled
triangles) with the dashed black line (open triangles)), again
confirming the evidence of an anomalous behavior in that period. In
2005, it appears that the growth rate relaxed back to the normal
level (according to the historical record).

\subsection{Yearly periodicity and intra-year structure with a scale
and translation
modulated model}

Inspired by these results, we propose the following quantitative
model. Consider a time $t$ in units of month. We write $t=12 T +m$,
where $T$ is the year and $m$ is the month within that year and thus
goes from $1$ (January) to $12$ (December). For
instance, $t = 26$ corresponds to $T=2$ and $m=2$ (February), while
$t=38$ corresponds to $T=3$ and again the same month $m=2$
(February) within the year. We propose to model the intra-year
structure of the growth rate $g(t)$ together with possible yearly
variations by the following expression \be
     g(t=12T+m) = f(T)h(m) + j(T) ~. \label{mvnnwsadam1}
\ee
In words, the growth rate has an intra-year structure $h(m)$
modulated from year to year in amplitude by $f(T)$
up to a possible overall translation $j(T)$ which can also vary
from year to year. We can expect
$f(T)$ and $j(T)$ to be approximately constant
for most years, except around 1990 and 2004 for which we should see
an anomaly in either or both of them, since these two periods
had bubbles. Note that this model (\ref{mvnnwsadam1})
gives an exact yearly periodicity if $f(T)$ and $j(T)$ are
constant. A non-constant $f(T)$ describes an amplitude modulation of
the yearly periodicity. In particular, we expect a strong peak
around $T=2004$. With this model, we can focus on predicting $f(T)$ and $j(T)$
only, because we have removed the complex intra-year structure.

We have thus fitted the model (\ref{mvnnwsadam1}) to three subsets
of the whole available time series for the growth rate $g(t)$ and
also to the whole set taken globally, in order to test for the
robustness of the model. For this, we use the cost function
\be
\sum_{T=1}^{T_{\rm max}} \sum_{m=1}^{12} [g(t=12T+m) - f(T)h(m) -
j(T)]^2
\ee
which is minimized with respect to the 12 unknown
variables $h(1), ..., h(12)$ and the $2 \times T_{\rm max}$
variables $[f(1), j(1)], ..., f(T_{\rm max}), j(T_{\rm max})$. There
are $12 T_{\rm max}$ terms in the sum and $12+ 2 \times T_{\rm max}$
unknown variables. This shows that the system is well-constrained
as soon as $T_{\rm max}\geq 2$.
For instance for $T_{\rm max}=20$, we have $52$ unknown variables to
fit and $240$ terms in the sum to constrain the fit.

\begin{figure}[htb]
\begin{center}
\includegraphics[width=8cm]{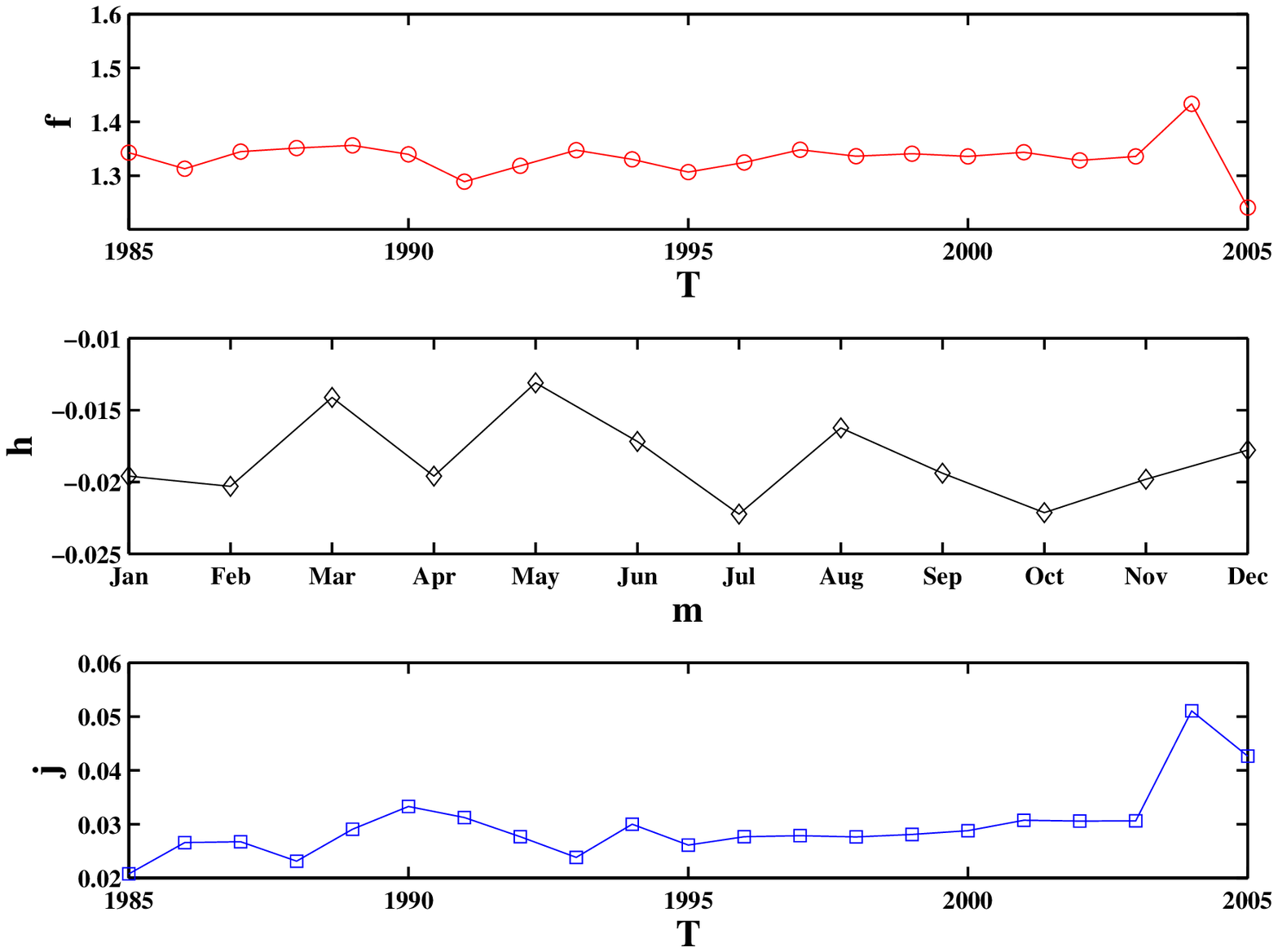}
\includegraphics[width=8cm]{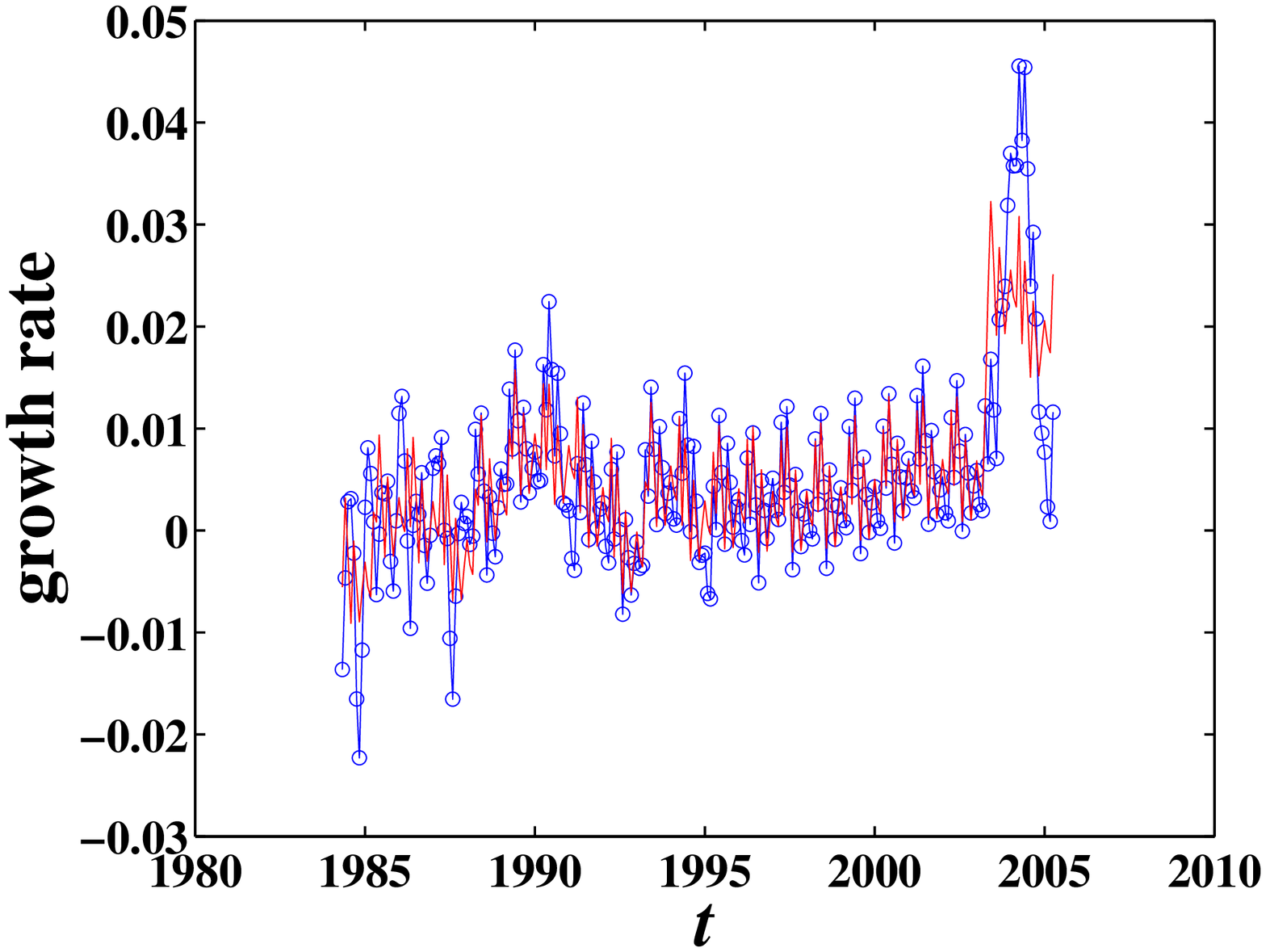}
\end{center}
\caption{Upper panels: three graphs showing the three functions
$f(T)$, $h(m)$ and $j(T)$ fitted on the growth rate over the whole
time interval from 1985 to 2005. Lower panel: Comparison between the
growth rate data (empty blue circles) and the model
(\ref{mvnnwsadam1}) (red line). } \label{FigModel10_fit}
\end{figure}

Figure \ref{FigModel10_fit} illustrates the result of the fit of
model (\ref{mvnnwsadam1}) to the growth rate over the whole time
interval from 1985 to 2005. As expected, we can observe a clear peak
in the amplitude $f(T)$ corresponding to the year 2004, while there
is not appreciable peak around 1990. This means that the recent
bubble appears significantly stronger than any other episodes in the
last 20 years and dwarfs them. The anomalous nature of the recent
bubble is reinforced by the existence of a peak in $j(T)$ for the
same year 2004, showing that both the amplitude and translation
components of the growth rates has been completely anomalous in
2004. The middle graph of the top panel of figure
\ref{FigModel10_fit} shows the intra-year pattern captured by the
model, which is in remarkable agreement with the pattern shown in
figure \ref{Fig:LV_Pred_g}: one can observe a peak in March, May,
August and December, the largest peak being in May. The bottom panel
of figure \ref{FigModel10_fit} shows visually how well (or badly)
the model fits the actual data. The quality of the fit is excellent,
except in 2004-2005. In other words, we clearly identify a very
anomalous or exceptional behavior in 2004-2005, again providing a
confirmation that something exceptional or anomalous has occurred
during that period.

\begin{figure}[htb]
\begin{center}
\includegraphics[width=6cm]{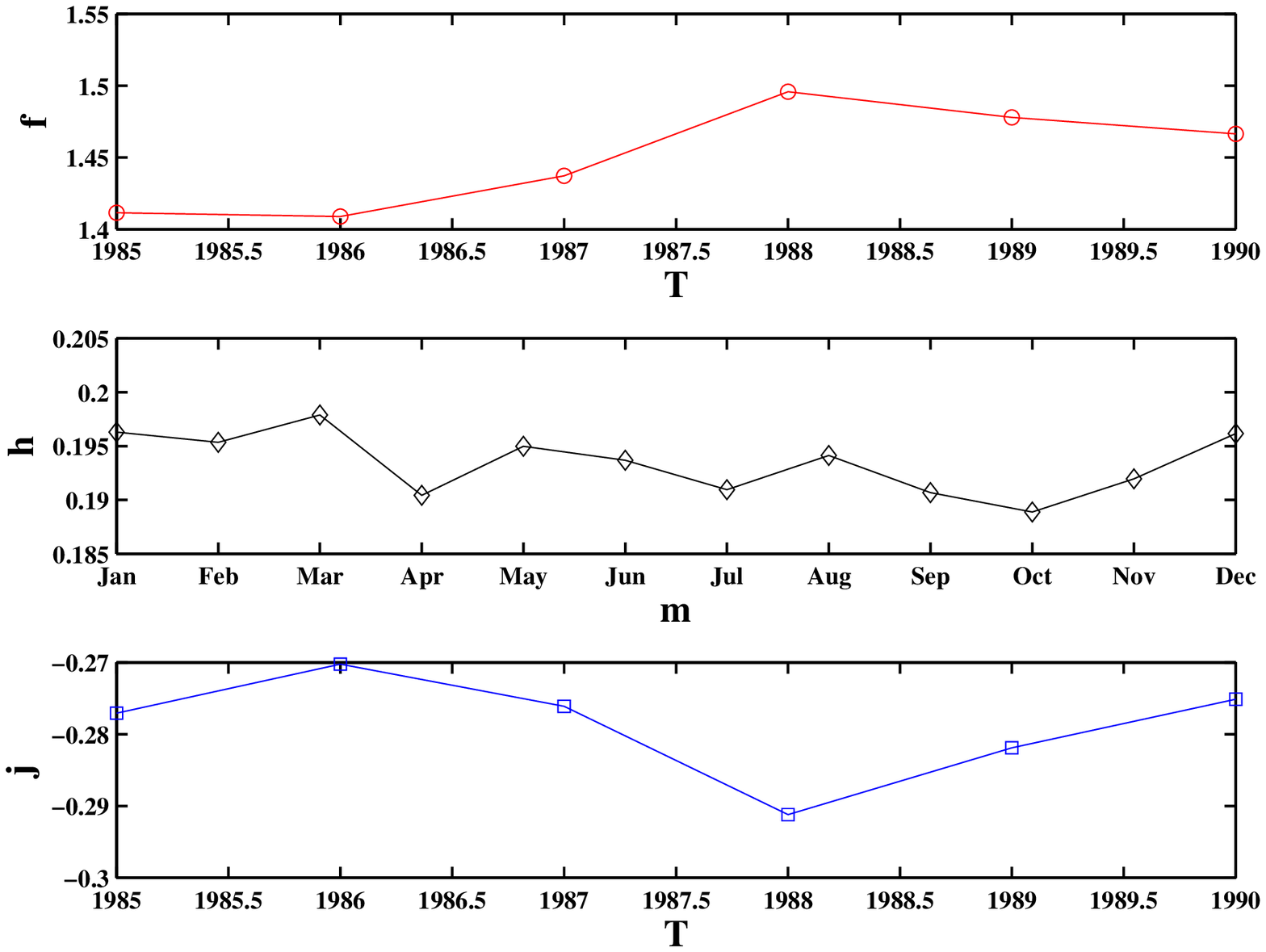}
\includegraphics[width=6cm]{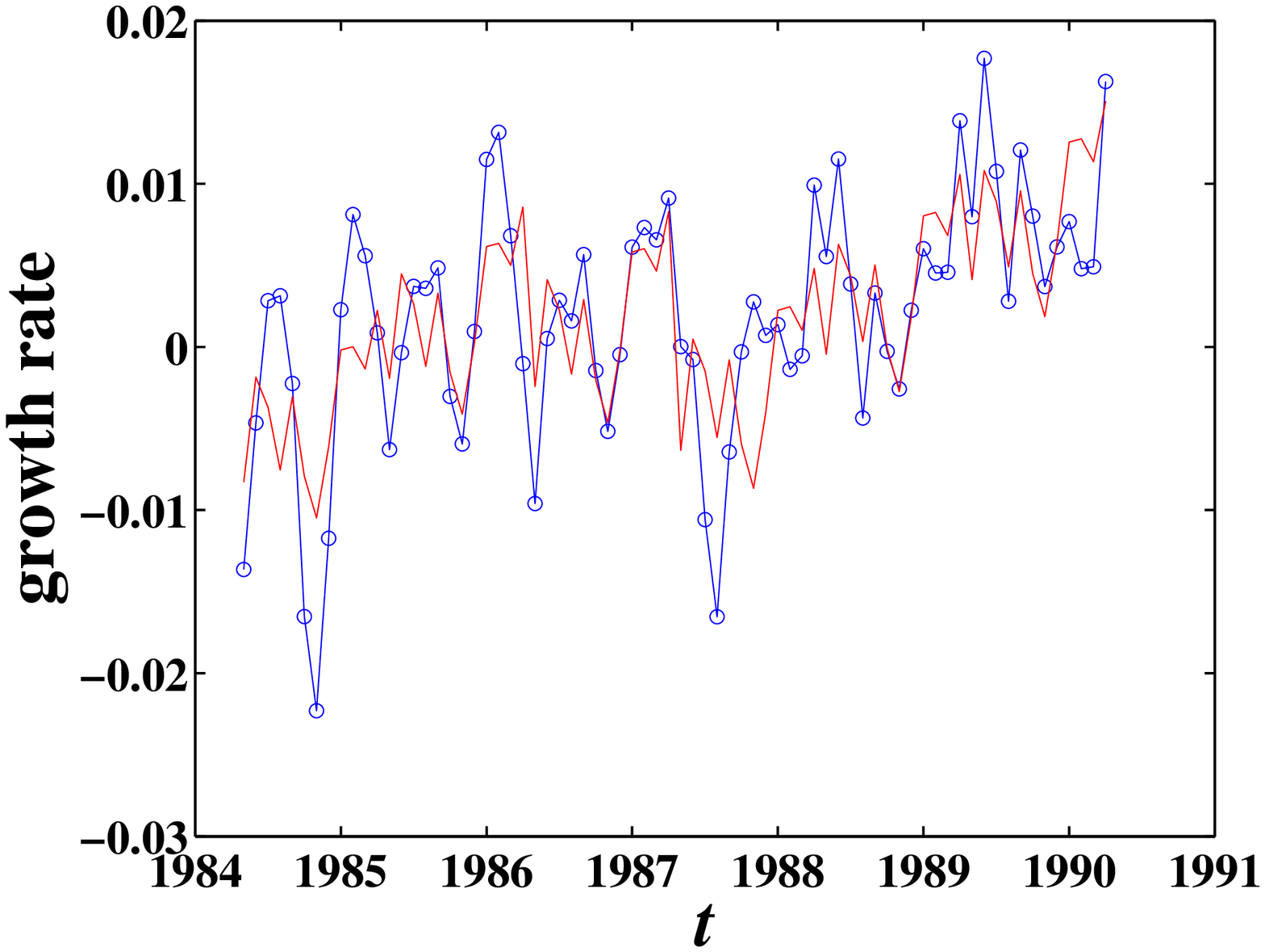}
\end{center}
\caption{Same as figure \ref{FigModel10_fit} for the period from
1985 to 1990. } \label{FigModel10_a}
\end{figure}

Figure \ref{FigModel10_a} is the same as figure \ref{FigModel10_fit} for the
period from 1985 to 1990. One can clearly here observe a peak in
the scaling amplitude $f(T)$ at $T=$1988 and in the translation
term $j(T)$ at $T=$1986, suggesting that the first bubble of the
1985-2005 period
occurred over a relatively large time period 1985-1990, with two
successive contributions. The intra-year structure $h(m)$ has also
its peaks on March, May, August and December, but this intra-year structure
is weaker than for other sub-periods. The lower panel of figure
\ref{FigModel10_a}
shows that the model captures very well the overall trend as well
as the intra-year structure. The main discrepancies are in the amplitude
of the large peaks and valleys, which are not fully predicted.

\begin{figure}[htb]
\begin{center}
\includegraphics[width=6cm]{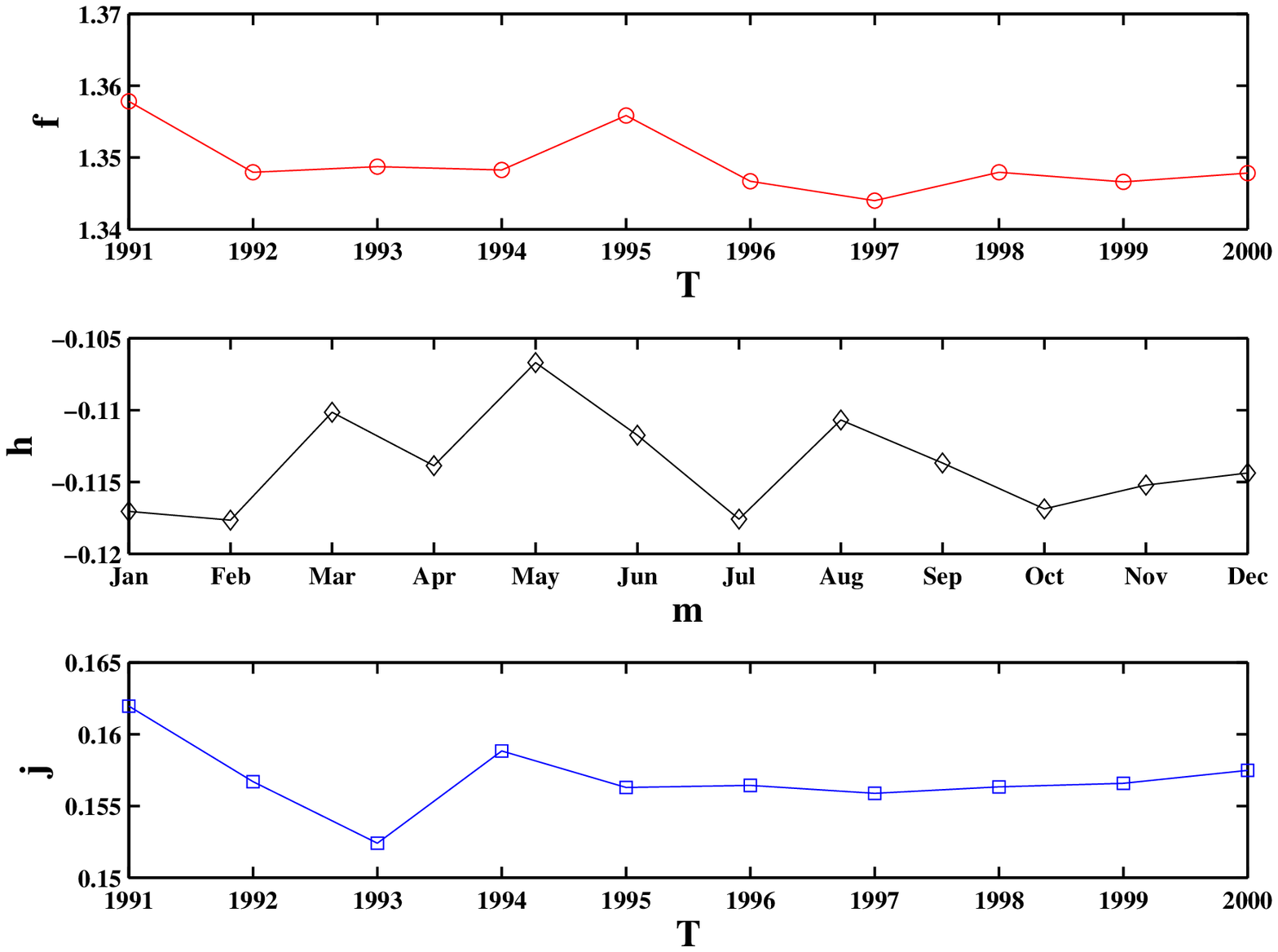}
\includegraphics[width=6cm]{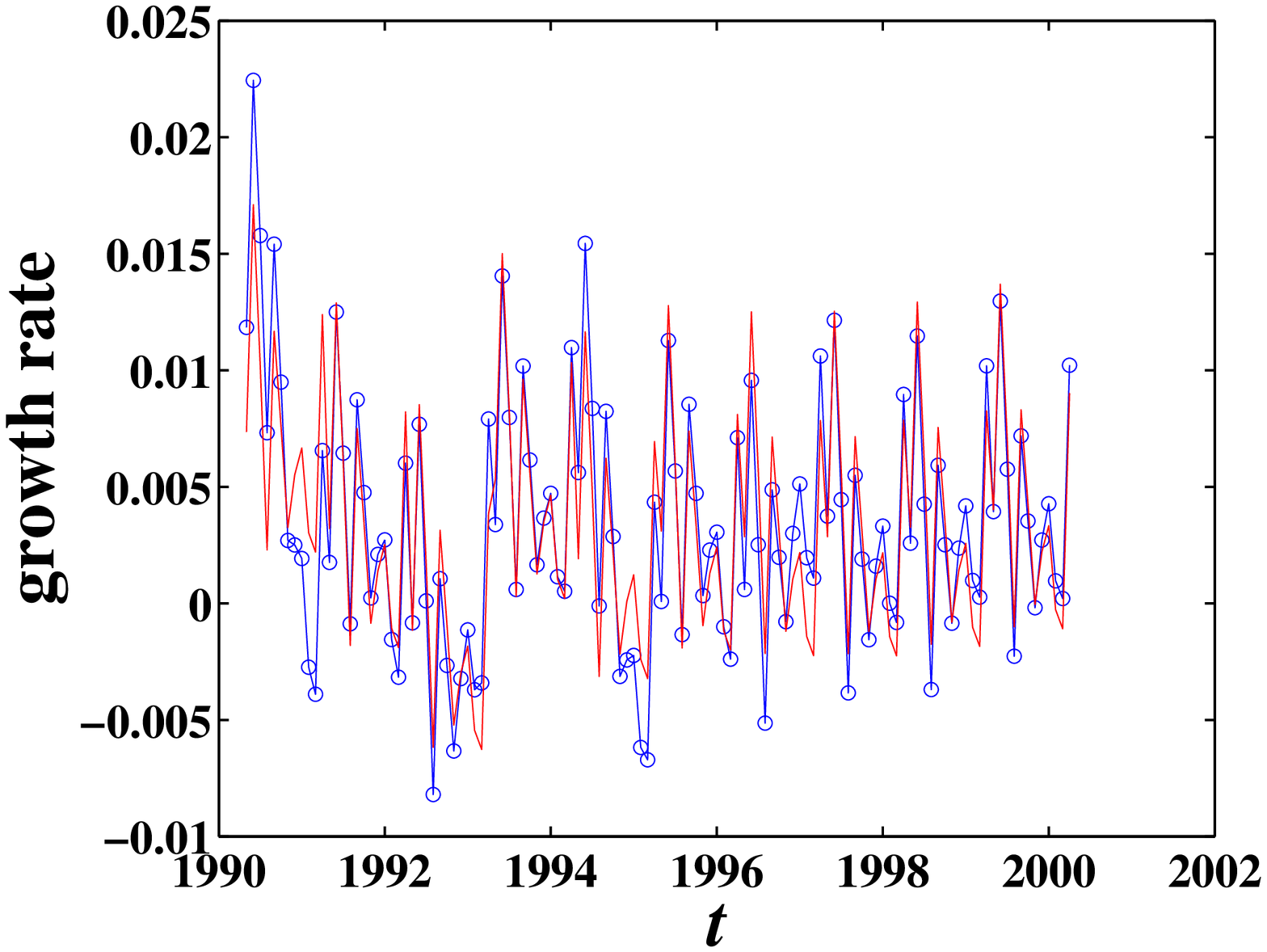}
\end{center}
\caption{Same as figure \ref{FigModel10_fit} for the period from
1991 to 2000. } \label{FigModel10_b}
\end{figure}

Figure \ref{FigModel10_b} is the same as figure \ref{FigModel10_fit}
for the period from 1991 to 2000. One can clearly here observe a
peak in the scaling amplitude $f(T)$ at $T=$1995 and in the
translation term $j(T)$ at $T=$1994. This thus identifies a small
bubble in the mid-1990s. The intra-year structure $h(m)$ has also
its peaks on March, May, August and December, with very large
amplitudes. The lower panel of figure \ref{FigModel10_b} shows a
truly excellent fit.

\begin{figure}[htb]
\begin{center}
\includegraphics[width=6cm]{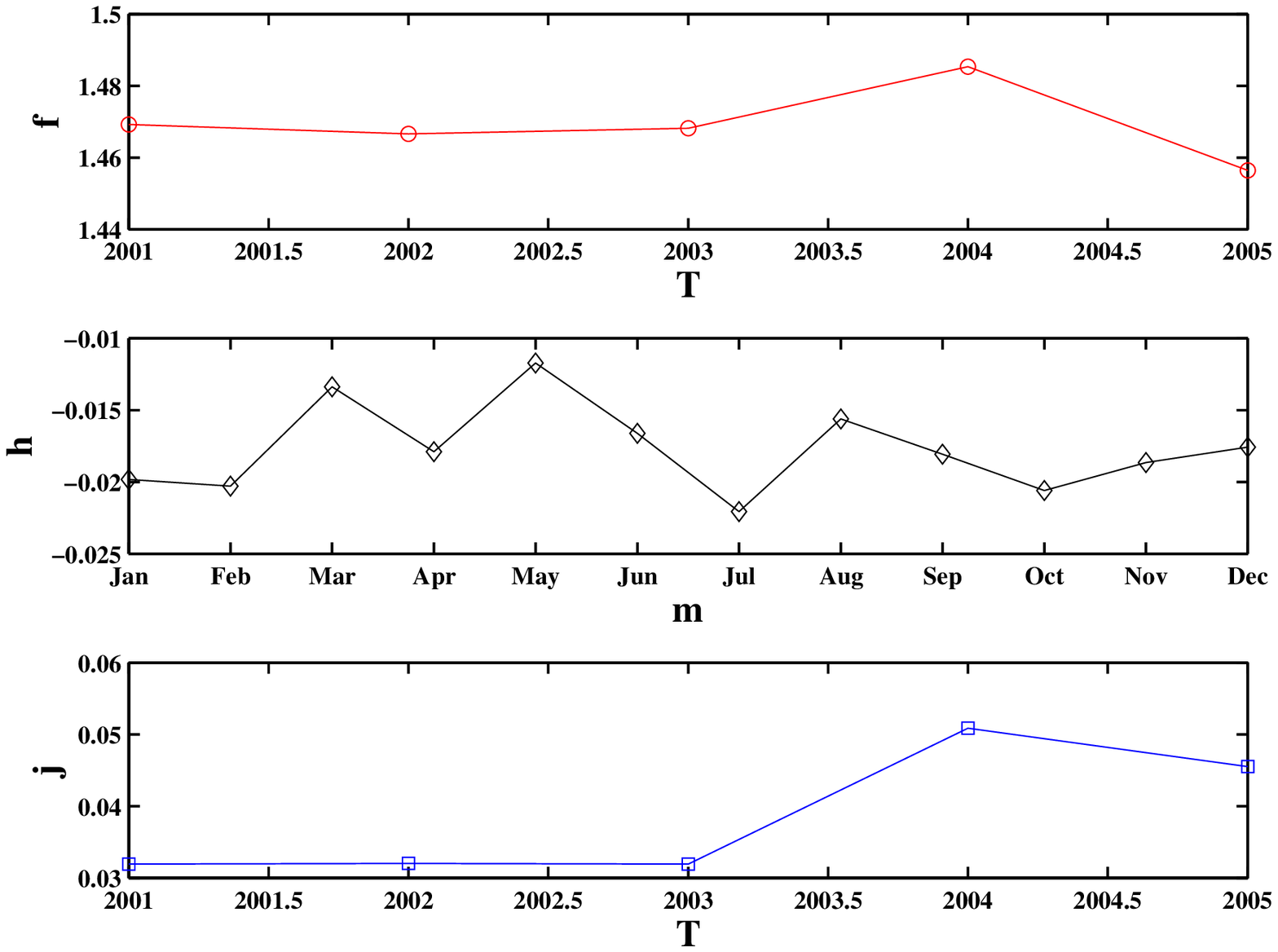}
\includegraphics[width=6cm]{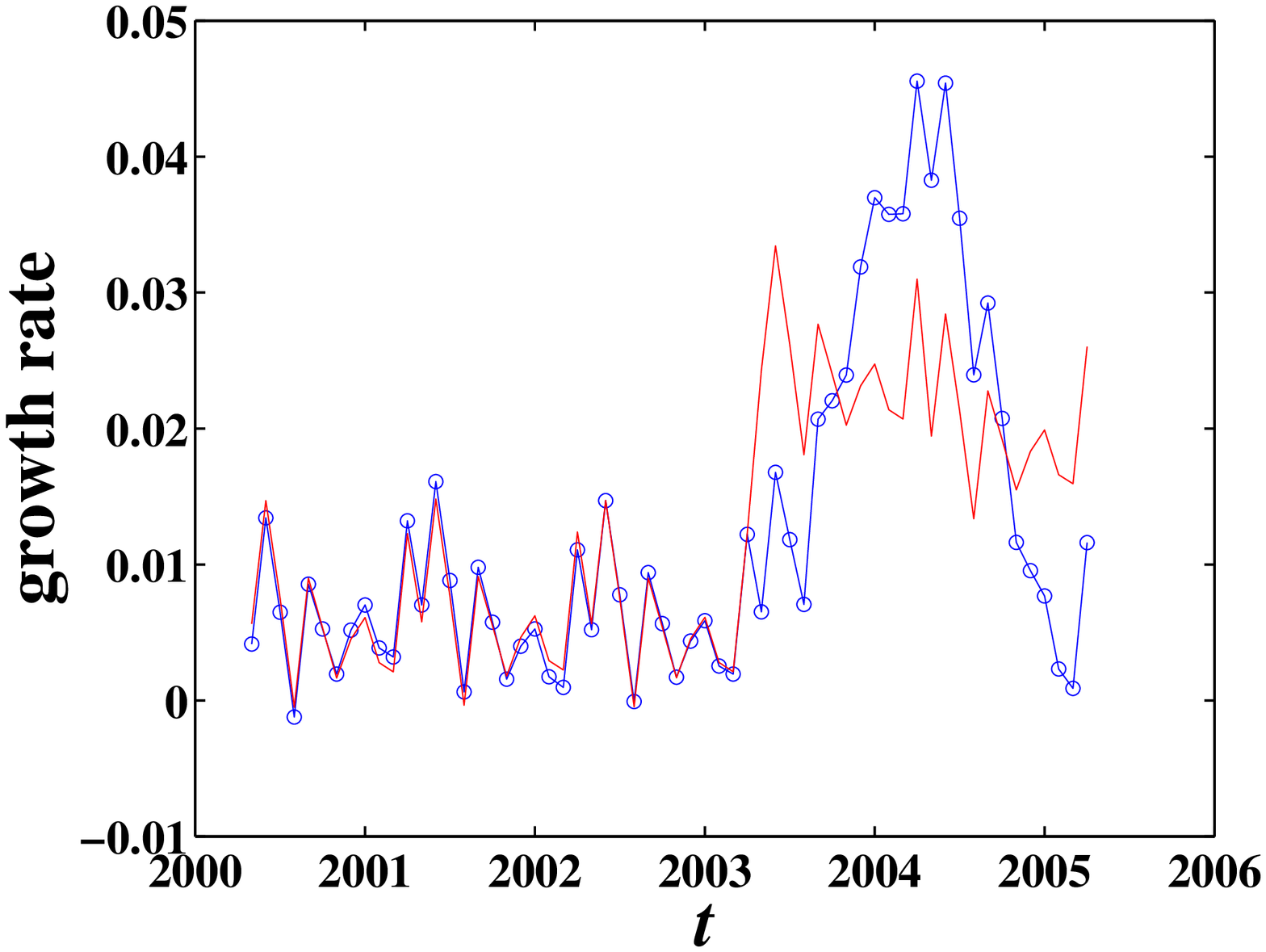}
\end{center}
\caption{Same as figure \ref{FigModel10_fit} for the period from
2001 to 2005. } \label{FigModel10_c}
\end{figure}

Figure \ref{FigModel10_c} is the same as figure \ref{FigModel10_fit} for the
period from 2001 to 2005. One can clearly here observe a peak in
the scaling amplitude $f(T)$ at $T=$2004 and in the translation
term $j(T)$ also at $T=$2004. This thus clearly identifies the bubble
as peaking in 2004.
The intra-year structure $h(m)$ has also
its peaks on March, May, August and December, with very large amplitudes
and very good agreement with the other three figures. The lower panel
of figure \ref{FigModel10_c} shows an excellent fit up to the
early 2003 and then
a rather large discrepancy starting early 2003 all the way to the last
data point approaching mid-2005. In particular, note that the intra-year
structure is washed out by the anomalous growth rate culminating in mid-2004.
Symmetrically, the intra-year structure is also absent in the fast
decay of the growth rate back to normal. We do not have enough data
to ascertain if the growth rate has resumed its normal intra-year pattern.
We believe that this is a very important diagnostic to characterize
abnormal behavior and this could be a very useful variable to
monitor on a monthly basis.

The four figures \ref{FigModel10_fit}-\ref{FigModel10_c} validate
model (\ref{mvnnwsadam1}): in particular, they show the very robust
intra-year structure with peaks in March, May, August and December.

One possible contribution to this quarterly periodicity comes from
the construction of the CSI: the monthly indexes use a three-month
moving average algorithm. Home sales pairs are accumulated in
rolling three-month periods, on which the repeat sales
 methodology is applied. The index point for each reporting month is based on
 sales pairs found for that month and the preceding two months. For example, the
December 2005 index point is based on repeat sales data for October,
November and December of 2005. This averaging methodology is used to
offset delays that can occur in the flow of sales price data from
county deed recorders and to keep  sample sizes large enough to
create meaningful price change averages. A three month rolling
window construction corresponds in general to a convolution of the
bare price with a kernel which possesses a three month periodicity
(or size). The Fourier transform of the convolution is the product
of Fourier transforms. Thus the spectrum of the signal should
contain the peaks of the Fourier spectrum of the kernel, which by
construction contains a peak at three months. However, our synthetic
tests (not shown) suggest that this effect is by far too small to
explain the strong amplitude of the observed quarterly periodicity.
It would be important to understanding why such intra-year structure
develops: is it the result of a natural intra-day organization of
buyers' behaviors associated with taxes/ income  constraints or a
problem of reporting or perhaps the effect of other calendar
regularities? Or is it the result of patterns coming from the supply
part of the equation, namely home-builders, developers, and perhaps
in the time modulation of the rates of allocated permits? Answering
these questions is important to determine how much emphasis one
should give to these results. But if indeed the intra-day structure
is a genuine non-artificial phenomenon, we believe that it offers a
remarkable opportunity for monitoring in real time the normal versus
abnormal evolution of the market and also for developing forecasts
on a month time horizon.

\subsection{Intra-year pattern from signs of growth rate increments}

The existence of a strong and robust intra-year structure in the
price growth rate can be further demonstrated by studying the sign
of $g(t+1)-g(t)$. A positive (negative) sign mean that the growth
rate tends to increase (decrease) from one month to the next.

Based on the seasonality of the growth rate, we are able to answer
the following question: given the current growth rate $g(t)$, will
the growth rate increase or decrease at time $t+1$? This amounts to
asking what is the sign of $g(t+1)-g(t)$? Technically,
we construct the
(unconditional) number of times the sign of the increment
$g(t+1)-g(t)$ is positive or negative irrespective of what is
$g(t)$. From Fig.~\ref{Fig:LV_CSW_gt}, we obtain a sequence of signs:
$--+-+--+--++$. For each month, we calculate the percentage of positive and
negative signs, respectively. The second and the third rows of Table
\ref{TB:sign} gives the percentage of positive and negative signs
for each month. The third and fourth rows gives the signs and the
associated percentages.

For instance, the table says that the ``probability'' of the sign of
$g(t={\rm{Feb}})-g(t={\rm{Jan}})$ being ``-'' is about 92.1\%. If we
know $g(t={\rm{Jan}})$, we can say that it is very probable that the
growth rate of February will be less than this January value. Thus,
this table has predictive power in the sense that the probabilities
to predict the signs are much higher than the value of $75\%$
obtained under the null hypothesis that $g(t)$ is a white noise process
\citep[see][]{Sornette-Andersen-2000-IJMPC}.
This table is another way to rephrase and expand on our preceding
analysis on the yearly periodicity by identifying a very strong and
robust intra-year structure.

\begin{table}[htb]
\caption{Analysis of the signs of $g(t+1)-g(t)$. The second and the third rows
gives the percentage of positive and negative signs
for each month. The third and fourth rows give the sign for each month that dominates
and the associated percentages.}
\begin{center}
\begin{tabular}{c|ccccccccccccccccccccccc}
       \hline\hline
       Mon & Jan & Feb & Mar & Apr & May & Jun & Jul & Aug & Sep & Oct
& Nov & Dec
       \\\hline
       +\% & 7.91&17.2&88.0&5.64&97.7&8.47&8.47&91.4&6.57&8.92&84.2&82.2\\
       -\% & 92.1&82.8&12.0&94.4&2.29&91.5&91.5&8.59&93.4&91.1&15.8&17.8\\
       sign&   - &  - &  + & -  & +  & -  & -  & + & - & - & + & + \\
       \%  &
       92.1&82.8&88.0&94.4&97.7&91.5&91.5&91.4&93.4&91.1&84.2&82.2\\
       \hline\hline
\end{tabular}
       \label{TB:sign}
\end{center}
\end{table}

Since our initial analysis performed in the summer of 2005 which used
data up to March 2005,
new data for the 27 CSW indexes has become available which
covers the interval from Apr. 2005 to
Sept. 2006. It is very interesting to check if the sign of the growth variations
obtained in Table \ref{TB:sign} using the data until March 2005 still
applies to the new data. The realized signs of
the newly available months are calculated and the sequence of signs
is the following: - (Apr. 2005, 27 CSW indexes out of 27), + (May. 2005, 27 out
of 27), - (Jun. 2005, 27 out of 27), - (Jul. 2005, 27 out of 27), +
(Aug. 2005, 27 out of 27), - (Sep. 2005, 27 out of 27), - (Oct.
2005, 27 out of 27), + (Nov. 2005, 27 out of 27), + (Dec. 2005, 21
out of 27), - (Jan. 2006, 27 out of 27), - (Feb. 2006, 27 out of
27), + (Mar. 2006, 27 out of 27), - (Apr. 2006, 27 out of 27), +
(May. 2006, 27 out of 27), - (Jun. 2006, 27 out of 27), - (Jul.
2006, 27 out of 27), + (Aug. 2006, 27 out of 27), and - (Sep. 2006,
27 out of 27). Thus,  table  \ref{TB:sign} predicts exactly the
signs of the growth rate variations of all 27 CSW indexes for all months
except for Dec. 2005 for which there are 6 errors: table  \ref{TB:sign}  predicts
that the growth rate variation from Dec. 2005 to Jan. 2006 should be +, which
is correct for 21 CSW indexes out of 27, corresponding to a success ratio of 77\% (close
to the white noise case).  This score is slightly lower than the previously estimated
probability of 82.2\% for the month of December, which is the lowest among all months. Overall, the
success rate is remarkably high, adding  further evidence that
the Las Vegas property market has returned to a more normal phase (no bubble
from April 2005 to Sept. 2006).

\section{Predicting the monthly growth rate}
\label{s1:prediction}

\begin{figure}[htb]
\begin{center}
\includegraphics[width=8cm]{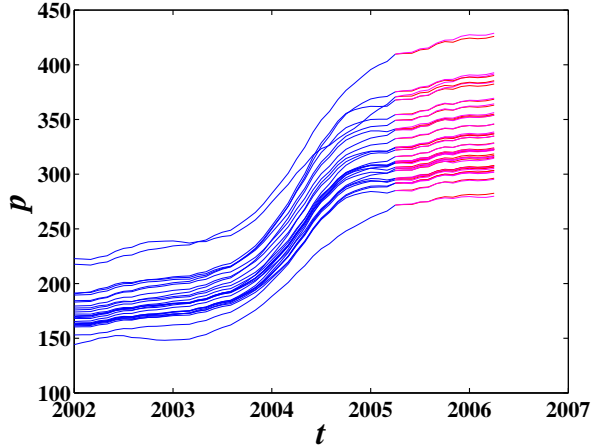}
\end{center}
\caption{Predicting regional CSW indexes one year ahead. Red lines:
Prediction using average growth rate obtained from all 27 indexes;
Magenta lines: Prediction using average growth rate obtained from
the individual index under investigation. The two kinds of prediction
are almost undistinguishable.}
\label{Fig:LV_Pred_CSW_Ps}
\end{figure}

Conditional of the evidence that the anomalous faster
than exponential growth has ended, let us attempt to
predict the future evolution of the CSW indexes based only
on the strong seasonality of the growth rate. Figure
\ref{Fig:LV_Pred_CSW_Ps} presents the predictions one year ahead for
the 27 regional CSW indexes. Two different prediction schemes are
used. The {\textcolor[rgb]{1.00,0.00,0.00}{RED}} lines are based on
the average growth rate obtained from all 27 indexes, while the
{\textcolor[rgb]{1.00,0.00,1.00}{MAGENTA}} lines are based on the
average growth rate obtained from the individual index under
investigation. There is not discernable difference.

\begin{figure}[htb]
\begin{center}
\includegraphics[width=8cm]{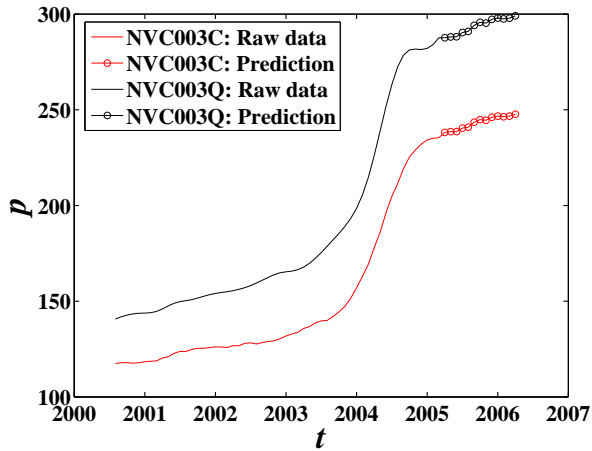}
\end{center}
\caption{Predicting Clark County (Las Vegas MSA) indexes (NVC003Q
and NVC003C) one year ahead.} \label{Fig:LV_Pred_CSW_NVs}
\end{figure}

A similar prediction of the Clark County (Las Vegas MSA) indexes
(NVC003Q and NVC003C) has also been made using the average growth
rates obtained from all 27 regional indexes. Since these two indexes
are only available from July-2000 to March-2005, we do not have
enough data to calculate the average growth rates using the indexes
themselves. The results are shown in Fig.~\ref{Fig:LV_Pred_CSW_NVs}.

\section{Conclusion}
\label{s1:concl}

We have analyzed 27 house price indexes of Las Vegas from Jun. 1983
to Mar. 2005, corresponding to 27 different zip codes. These
analyses confirm the existence of a real-estate bubble, defined as a
price acceleration faster than exponential. This bubble is found
however to be confined to a rather limited time interval in the
recent past from approximately 2003 to mid-2004 and has
progressively transformed into a more normal growth rate in 2005.
The data up to mid-2005 suggests that the current growth rate has
now come back to pre-bubble levels. We conclude that there has been
no bubble from 1990 to 2002 except for a medium-sized surge in 1995,
then a short-lived but very strong bubble until mid-2004 which has
been followed by a smoothed transition back to what appears to be
normal. It thus seems that, while the strength of the real-estate
bubble has been very strong over the period 2003-2004, the price
appreciation rate has returned basically to normal.

In addition, we have identified a strong yearly periodicity which
provides a good potential for fine-tuned prediction from month to
month. As the intra-year structure is likely a genuine
non-artificial phenomenon, it offers a remarkable opportunity for
monitoring in real time the normal versus abnormal evolution of the
market and also for developing forecasts on a monthly time horizon.
In particular, a monthly monitoring using a model that we have
developed here could confirm, by testing the intra-year structure,
if indeed the market has returned to ``normal'' or if more
turbulence is expected ahead. In addition, it would provide a
real-time observatory of upsurges and other anomalous behavior at
the monthly scale. This requires additional technical developments
and tests beyond this report.

Compared with previous analysis of
\citet{Zhou-Sornette-2003a-PA,Zhou-Sornette-2006b-PA} at the scale
of states and whole regions (northeast, midwest, south and west),
the present analysis demonstrates the existence of very significant
variations at the local scale, in the sense that the bubble in Las
Vegas seems to have preceded the more global USA bubble and has
ended approximately two years earlier (mid 2004 for Las Vegas
compared with mid-2006 for the whole of the USA).

%\bigskip
%{\textbf{Acknowledgments:}}
%
%We are grateful to Signature Homes (Las Vegas) for providing us the
%data. This work was partially supported by NSFC (Grant 70501011),
%Fok Ying Tong Education Foundation (Grant 101086), and the Alfred
%Kastler Foundation which supported W.-X. Zhou for a visiting
%position in France.

%\cite{testcite}

\bibliography{E:/papers/Auxiliary/Bibliography_FullJournal}

%\bibliography{Bibliography}

{\bf{Biographies}}:

{\bf{Wei-Xing Zhou}} is a Professor of Finance at the School of
Business in the East China University of Science and Technology. He
received his PhD in Chemical Engineering from the East China
University of Science and Technology in 2001. His current research
interest focuses on the modeling and prediction of catastrophic
events in complex systems.

{\bf{Didier Sornette}} holds the Chair of Entrepreneurial Risks at
the Department of Management, Technology and Economics of ETH
Zurich. He received his PhD in Statistical Physics from the
University of Nice, France. His current research focuses on the
modeling and prediction of catastrophic events in complex systems,
with applications to finance, economics, seismology, geophysics and
biology.

\end{document}